# Magnetic Circular Dichroism in Hyperbolic Metamaterial Nanoparticles

Joel Kuttruff[1,2]¶, Alessio Gabbani[3]¶, Gaia Petrucci[3], Yingqi Zhao[4], Marzia Iarossi[4,5], Esteban Pedrueza-Villalmanzo[6], Alexandre Dmitriev[6], Antonietta Parracino[7], Giuseppe Strangi[8,9], Francesco De Angelis[4], Daniele Brida[1], Francesco Pineider[3]*, and Nicolò Maccaferri[1]*

[1]Department of Physics and Materials Science, University of Luxembourg, 162a avenue de la Faincerie, 1511, Luxembourg, Luxembourg
[2]Department of Physics, University of Konstanz, Universitaetsstrasse 10, 78464 Konstanz, Germany
[3]Dipartimento di Chimica e Chimica Industriale, Università di Pisa, Via Giuseppe Moruzzi 13, 56124, Pisa, Italy
[4]Plasmon Nanotechnologies Unit, Istituto Italiano di Tecnologia, Via Morego 30, 16163, Genova, Italy
[5] Dipartimento di Informatica, Bioingegneria, Robotica e Ingegneria dei Sistemi (DIBRIS). Università degli Studi di Genova, Via Balbi 5, 16126 Genova, Italy
[6]Department of Physics, University of Gothenburg, Universitetsplatsen 1, 405 30, Gothenburg, Sweden
[7]Department of Chemistry, Uppsala University, Husargatan 3, 752 37, Uppsala, Sweden
[8]Department of Physics, Case Western Reserve University, 10600 Euclid Avenue, 44106, Cleveland, Ohio, USA
[9]CNR-NANOTEC Istituto di Nanotecnologia and Department of Physics, University of Calabria, Via Pietro Bucci 87036, Rende, Italy
*francesco.pineider@unipi.it
*nicolo.maccaferri@uni.lu
¶contributed equally

**The optical properties of some nanomaterials can be controlled by an external magnetic field, providing active functionalities for a wide range of applications, from single-molecule sensing to nanoscale nonreciprocal optical isolation. Materials with broadband tunable magneto-optical response are therefore highly desired for various components in next-generation integrated photonic nanodevices. Concurrently, hyperbolic metamaterials received a lot of attention in the past decade since they exhibit unusual properties that are rarely observed in nature and provide an ideal platform to control the optical response at the nanoscale via careful design of the effective permittivity tensor, surpassing the possibilities of conventional systems. Here, we experimentally study magnetic circular dichroism in a metasurface made of type-II hyperbolic nanoparticles on a transparent substrate. Numerical simulations confirm the experimental findings, and an analytical model is established to explain the physical origin of the observed magneto-optical effects, which can be described in terms of the coupling of fundamental electric and magnetic dipole modes with an external magnetic field. Our system paves the way for the development of nanophotonic active devices combining the benefits of sub-wavelength light manipulation in hyperbolic metamaterials supporting a large density of optical states with the ability to freely tune the magneto-optical response via control over the anisotropic permittivity of the system.**

Artificial anisotropic metamaterials made of metal-dielectric multilayers have recently emerged as a versatile platform to engineer light-matter interaction at the nanoscale [1]. Initially motivated by their ability to exhibit negative refraction and index [2], applications have diversified towards sensing [3,4] nanoscale waveguiding [5,6], enhanced nonlinearities [7-10] and amplified spontaneous emission [11-14]. The bulk optical properties of these materials can be described by an effective dielectric tensor, where the in-plane and out-of-plane permittivities, $\varepsilon_\parallel$ and $\varepsilon_\perp$, respectively, can be almost arbitrarily tuned by the geometric dimensions and material composition [15]. By stacking multiple layers of metal and dielectric materials, extreme optical anisotropies are achieved, often described by hyperbolic isofrequency surfaces ($\varepsilon_\parallel \cdot \varepsilon_\perp < 0$), which are opposed to the elliptical ones usually observed in typical materials. In particular, these hyperbolic metamaterials are distinguished as type-I ($\varepsilon_\parallel > 0, \varepsilon_\perp < 0$) and type-II ($\varepsilon_\parallel < 0, \varepsilon_\perp > 0$), depending on the sign of the permittivities along the in-plane and out-of-plane directions. This optical anisotropy can be further exploited to achieve enhanced magneto-optical response of a material [16], allowing to control the optical properties by applying an external magnetic field, which is fundamental in view of the development of active magneto-photonic devices [17-22], from sensing [23-27] and all-optical switching [28,29], to enhanced nonreciprocal light propagation [30-33]. Magneto-optical effects in type-I artificial hyperbolic metamaterials based on Au-Ni nanorod arrays have been recently reported [34]. As well, the potential to magnetically modulate the radiative heat transfer by using hyperbolic $InSb/SiO_2$ multilayers has been recently predicted [35], paving the way for exciting research directions enabled by the active functionalities provided by hyperbolic magneto-photonic nanomaterials. Recently, it was also shown that type-II hyperbolic metamaterial nanoparticles display a strong optical anisotropy in the visible and near-infrared spectral ranges, where separation between radiative and non-radiative spectral responses allows for on-demand control of absorption and scattering of light at the nanoscale [36,37].

Here, we focus on the interaction of an external magnetic field with the localized electromagnetic electric and magnetic modes excited in type-II hyperbolic metamaterial nanoparticles. We first demonstrate that the optical properties of this system can be perfectly described by an anisotropic homogenized permittivity in the framework of an effective medium theory, thus proving their hyperbolic nature. The broadband magneto-optical response of the sample is then studied by magnetic circular dichroism (MCD) spectroscopy. We observe a lifting of the degeneracy of circular magneto-plasmonic modes at both fundamental resonances of the system. Based on numerical modelling of the structures, we interpret the experimentally measured MCD in terms of magnetic field-induced spatial confinement/broadening of circular currents in the nanoparticles. Finally, by using a simple analytical model system in the framework of Mie theory, we show that the magneto-optical response of the system can be described by considering the coupling of electric and magnetic dipole resonances to the external magnetic field. The system presented here allows the development of nanophotonic active devices combining the benefits of sub-wavelength light manipulation with the ability to freely tune the magneto-optical response via control over the anisotropy in the permittivity of the system, potentially at the single nanoparticle level.

Disk-shaped hyperbolic nanoparticles with a nominal diameter of 250 nm were produced on fused silica substrates using hole-mask colloidal lithography (for more details on the fabrication see the **Methods** section) **[38]**. Two types of hybrid nanostructures were prepared, in which the dielectric material is varied between titania ($TiO_2$) and silica ($SiO_2$). By changing the dielectric material, a tuning of the spectral position of the fundamental resonance is possible **[36]**. For both the Au/$TiO_2$ and Au/$SiO_2$ hybrid nanoparticles, five alternating layers of Au (10 nm each) and of the dielectric material (20 nm each) were deposited, leading to a total height of 150 nm (see inset of **Fig. 1a** for a sketch of the sample design). A scanning electron microscopy (SEM) image of the Au/$TiO_2$ sample is shown in the inset of **Fig. 1a**, alongside with the experimentally obtained optical extinction (1 – T, T being the transmittance of the metasurface). For the Au/$TiO_2$ two peaks are clearly visible in the extinction spectrum, at $\lambda \approx 1000$ nm and $\lambda \approx 1600$ nm. This unusual optical response originates in the hyperbolic dispersion of the multilayer dielectric function. Indeed, the optical properties of these nanoparticles can be described by an effective medium theory, where the anisotropy is represented by an in-plane permittivity $\varepsilon_\parallel$ and out-of-plane permittivity $\varepsilon_\perp$. These permittivities can be expressed using the following well-known relations **[15]**

$$\varepsilon_\parallel = \frac{t_m \varepsilon_m + t_d \varepsilon_d}{t_m + t_d} \qquad (1)$$

$$\varepsilon_\perp = \frac{\varepsilon_m \varepsilon_d (t_m + t_d)}{t_m \varepsilon_d + t_d \varepsilon_m}, \qquad (2)$$

where $t_{m/d}$ is the thickness and $\varepsilon_{m/d}$ the permittivity of the metal/dielectric. The dispersion of the permittivity is shown as a function of the wavelength in the inset of **Fig. 1b**. Above 500 nm, $\varepsilon_\perp$ is positive (dielectric-like response) while $\varepsilon_\parallel$ is negative (metal-like response), corresponding to a type-II hyperbolic dispersion relation. To validate our approach, numerical simulations have been performed based on the finite element method (FEM, details on the simulations can be found in the **Methods** section). The optical extinction was calculated considering, firstly, the real structure with alternating layers (solid line in **Fig. 1b**) and, secondly, a homogenized structure described by an anisotropic effective permittivity as expressed in **Eqs. (1)** and **(2)** (dashed line in **Fig. 1b**). The resulting spectra are in excellent agreement and describe well the extinction experimentally measured. It is worth noting that the high energy plasmon decays mainly radiatively, while the low energy plasmon decays mainly non-radiatively, as can be seen in **Supplementary Fig. 1** and in Ref. **[36]**.

The ability to separately tune both decay channels contrasts with conventional plasmonic architectures (cf. plain gold nano-disks, **Supplementary Fig. 2a**) and clearly originates from the hyperbolicity of the permittivity of our system. Having assessed the hyperbolic nature of these nanoparticles, we now move to study their magneto-optical response.

When a magnetic field is applied to a material, the moving charge carriers undergo a Lorentz force $\mathbf{F_L}$, oriented perpendicularly to the magnetic field and the trajectory of the charges, i.e.

$$\mathbf{F_L} = q(\mathbf{v} \times \mathbf{B}). \quad (3)$$

In this case when a static field H is applied along, for instance, the z-direction (out-of-plane), the symmetry of the dielectric function of the material is broken, leading to non-zero off-diagonal components of the permittivity [39], i.e. the tensorial permittivity $\hat{\varepsilon}$ generalizes to

$$\hat{\varepsilon} = \begin{pmatrix} \varepsilon_{xx} & \varepsilon_{MO} & 0 \\ -\varepsilon_{MO} & \varepsilon_{yy} & 0 \\ 0 & 0 & \varepsilon_{zz} \end{pmatrix} \quad (4)$$

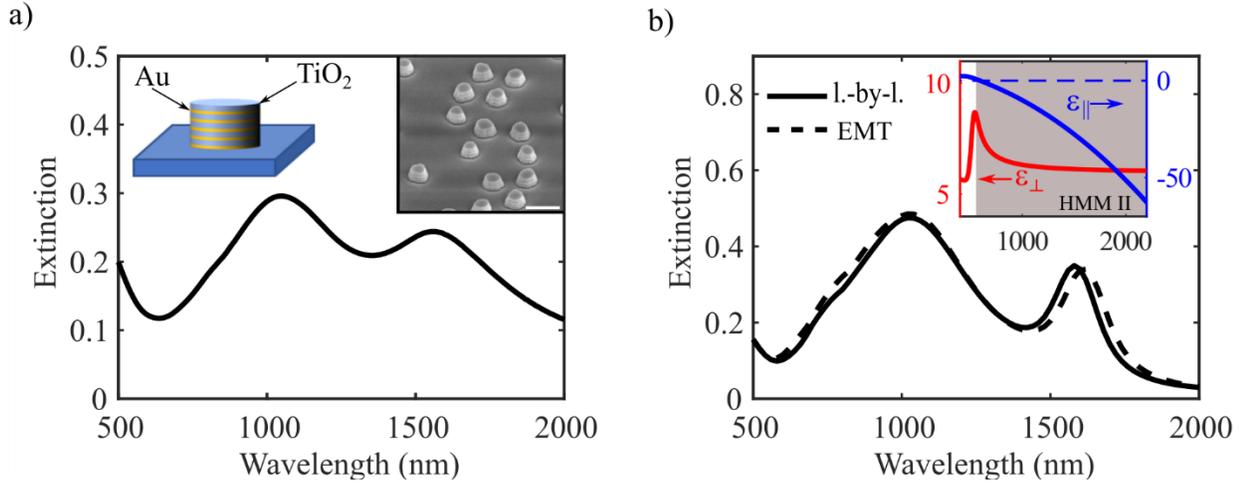

**FIG. 1. Optical extinction of hyperbolic nanoparticles.** a) Experimental extinction spectrum of the Au/TiO$_2$ hyperbolic nanoparticles metasurface. Inset are a sketch of the geometry (top left) as well as a scanning electron microscopy (SEM) image of the sample (top right, scale bar is 400 nm). b) Numerically obtained extinction spectra calculated based on the real structure (layer-by-layer, solid line) and using an effective medium theory (EMT, dashed line). Inset are the real parts of the in-plane (blue) and out-of-plane (red) permittivities used in the EMT approach. Above 500 nm, the system shows type-II hyperbolic dispersion (HMM II), as indicated in the figure.

The non-diagonal magneto-optical (MO) term $\varepsilon_{MO}$ is generally very small for noble metals compared to ferromagnetic materials. Here, we assume $\varepsilon_{MO} = 10^{-4} + i10^{-3}$, valid for Au at 1 T, whereas $\varepsilon_{MO} = 1 + i0.1$ for ferromagnets like Co [39]. However, nanoscale design of bulk materials can drastically alter the magneto-optical properties due to the presence of plasmon resonances, leading to an enhanced magneto-optical response [39,40]. The magneto-plasmonic coupling is particularly interesting in the context of circular modes, where dichroism is induced by the applied magnetic field [21], directly prompting active modulation of the plasmon resonances using magnetic fields with applications in optical switching and refractometric sensing. Such circular modes can be excited by circularly polarized light in systems that otherwise exhibit rotational symmetry in the propagation direction of the incoming light. When a static magnetic field is applied along the propagation direction, the degeneracy of left- and right-handed modes is lifted due to the off-diagonal coupling term in **Eq. 4**, which is observable by a derivative-like spectral shape of the circular dichroism signal. In conventional plasmonic nanomaterials, such a magnetically induced mode splitting is observed at the fundamental electric dipole resonance (as shown for plain gold nano-disks in **Supplementary Fig. 2b**) and already studied in colloidal particles, for instance in Refs. [21] and [22]. Using a simple analytical model, we will show below that in addition to an electric dipole mode, the low energy resonance of the hyperbolic system can be understood as a fundamental magnetic resonance induced by the anisotropy in the dielectric function. Here, by considering the coupling of the charges with the magnetic field, we predict circular dichroism also at this low energy magnetic dipole mode of the hyperbolic nanoparticles, thereby pushing forward previous studies purely based on electric dipole resonances. We numerically assess the induced changes in the current density inside a single nanoparticle due to the interaction with an external magnetic field. All quantities are evaluated inside the topmost gold layer of the structure, which is resonantly excited at the low energy magnetic dipole mode. For this case, the numerically calculated current density **j** at vanishing magnetic field (H=0) is shown as white arrows in **Fig. 2a** for one phase of the incoming electric field together with the absolute value of the current density j (color coded). When a static magnetic field is applied in positive or negative z-direction (+H and –H, respectively), we expect a magnetically induced change in the current, due to the magneto-optical coupling **(Eq. 4)** in the gold layers, i.e.

$$\boldsymbol{j}(+H) = \boldsymbol{j}(0) + \Delta \boldsymbol{j}_{MO} \quad (5)$$

$$\boldsymbol{j}(-H) = \boldsymbol{j}(0) - \Delta \boldsymbol{j}_{MO}. \quad (6)$$

From **Eqs. (5)** and **(6)**, $\Delta \boldsymbol{j}$ can thus be calculated as $2\Delta \boldsymbol{j}_{MO} = \boldsymbol{j}(+H) - \boldsymbol{j}(-H)$.

Here, the current densities $j(+H)$ and $j(-H)$ are numerically obtained from separate simulations, where the direction of the applied magnetic field is inverted along the out-of-plane direction (see sketch in **Fig. 2a**). In addition, from $j(0)$ we can also directly compute the Lorentz force acting on the charge carriers as the cross product of the current density and the magnetic field. Indeed, the change in current density $\Delta j_{MO}$ follows well the direction of this force, as can be seen by comparing the arrows, indicating the direction of current flow, in **Fig. 2b** and **Fig. 2c**. One can further notice orientation deviations of the arrows at the boundary of the particle, which is expected from the shape anisotropy of the symmetric system at its boundaries. It should be noted at this point that the inversion of the magnetic field direction is topologically equivalent to a change in the helicity of the incoming light. Thus, for a fixed magnetic field and if averaged over one period of the electric field oscillation, following the sketch presented in **Fig. 2d**, the magnetically induced force on the free charge carriers will lead to spatial confinement or broadening of the circular modes, strictly depending on the helicity of the incoming electric field. This is verified via the change in absolute (i.e., independent on the phase $\varphi$) current density Δj in **Fig. 2e** and **Fig. 2f**, and consequently results in a shift of the resonance frequency for opposite helicities (i.e., circular dichroism).

To experimentally probe the interaction of the circular modes with an external magnetic field, we use MCD spectroscopy, which has been proved to be a promising tool to elucidate the symmetry of plasmonic modes, as well as to actively modulate the plasmon resonance **[41]**. We define the MCD via the difference between two extinction spectra acquired using light with opposite helicity, $E^+$ and $E^-$ respectively, in the presence of an external magnetic field parallel to the incident light direction,

$$\text{MCD} = \frac{E^+ - E^-}{E^+ + E^-}. \qquad (7)$$

Corresponding experimental MCD spectra for the Au/TiO$_2$ (red diamonds) and Au/SiO$_2$ (yellow circles) samples are shown in **Fig. 3a**. A derivative-like spectral shape is clearly observed at the low and the high energy resonances for the Au/TiO$_2$ sample. Such line-shape is consistent with the excitation of circular plasmonic modes in noble metal nanostructures with rotational symmetry in the plane of the incident electric field, as we show for plain gold nano-disks in **Supplementary Fig. 2.** For the hyperbolic nanoparticles, the same line shape of the MCD spectrum is observed for both the electric and magnetic dipole modes, in agreement with the rationalization reported above, i.e. magnetically induced spatial confinement/broadening of circular currents in the nanoparticles upon circularly polarized light excitation (**Fig. 2**). Interchanging the dielectric layer with SiO$_2$ then leads to a strong blueshift of the low-energy resonance, shown by the yellow circles in **Fig. 3a** (see also the linear extinction in **Supplementary Fig. 1**).

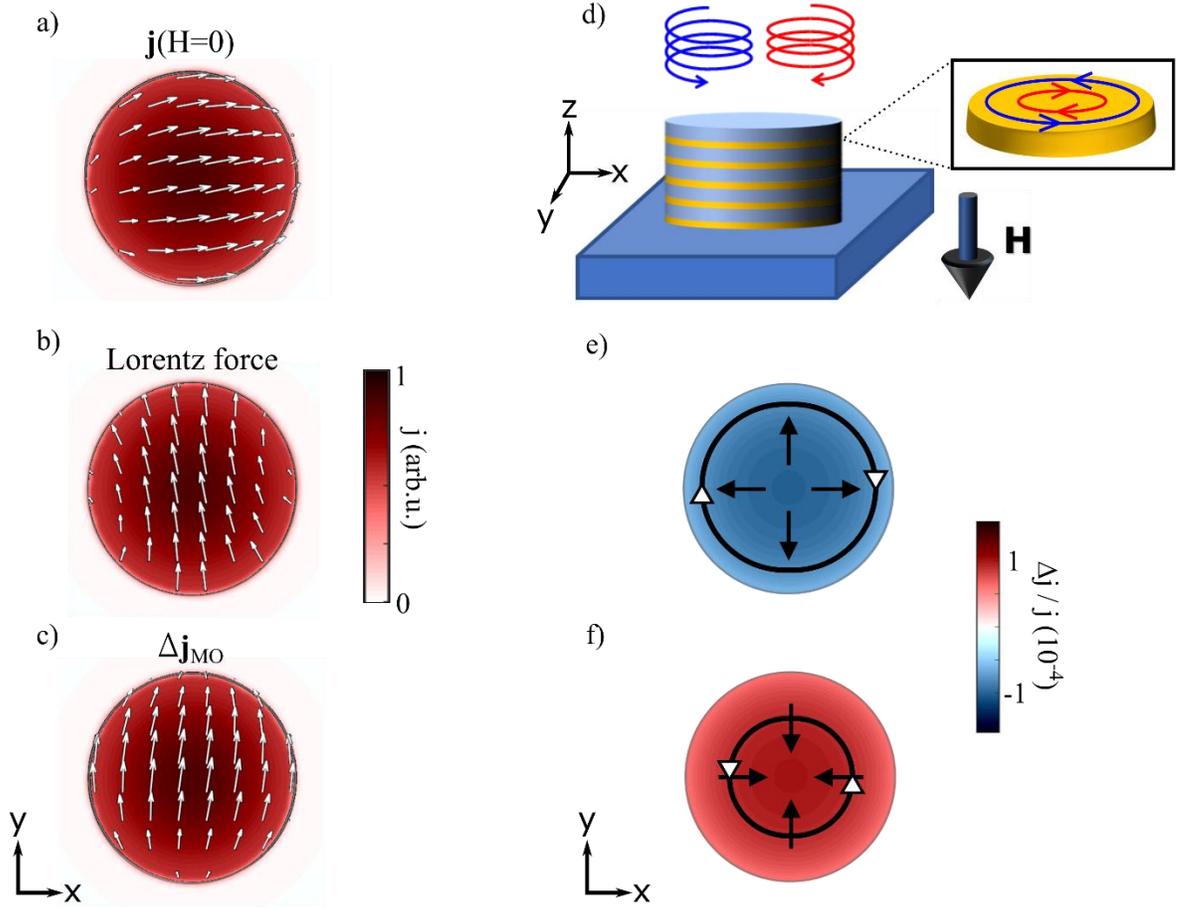

**Fig. 2. Circular currents in hyperbolic nanoparticles.** a)-c) Calculated current density **j** (white arrows) at the low energy resonance for zero magnetic field (a), Lorentz force calculated as cross product of **j** and the unit vector in direction of **H** (b), and magnetically induced change of current $\Delta \mathbf{j}_{MO} = 1/2(\mathbf{j}(-H) - \mathbf{j}(+H))$ (c). The absolute value of j is color coded for all three cases. d) Sketch of the physical domain displaying the circular dichroism effect in terms of spatial confinement/broadening of circular modes. e) & f) Absolute value of the induced relative change of current density Δj/j (color coded) for opposite helicity of the incoming electric field, as sketched in the plot. Black arrows show the gradient of Δj/j and serve as a guide to the eye displaying the induced spatial broadening/confinement of the circular plasmonic motion, thereby lifting the degeneracy of the modes.

To describe the magneto-optical properties of our system in the framework of an effective medium theory, **Eqs. (1)** and **(2)** are generalized as

$$\hat{\varepsilon}_{\mathrm{EMT}} = \begin{pmatrix} \varepsilon_\parallel & f * \varepsilon_{MO} & 0 \\ -f * \varepsilon_{MO} & \varepsilon_\parallel & 0 \\ 0 & 0 & \varepsilon_\perp \end{pmatrix}, \quad (8)$$

where the off-diagonal component due to the coupling with the external magnetic field is considered for gold, and $f = 1/3$ is the volume fraction of gold in the nanoparticle. Separate simulations are performed for both helicities and the MCD is calculated based on the individual extinction spectra according to **Eq. (7)**. Resulting spectra are shown in **Fig. 3b** and reproduce well the observed features, i.e., derivative-like spectral shape at both the high and low energy resonances of the hyperbolic nanoparticles. As already mentioned above, we assumed a constant value of the magneto-optical coupling term $\varepsilon_{MO} = 10^{-4} + i10^{-3}$. Sepúlveda et al. **[39]** further suggested a magneto-optical coupling term for the noble metals that is linear in the diagonal permittivity of the material, $\varepsilon_{MO} \propto \varepsilon_{xx}$. Considering such a frequency dependent $\varepsilon_{MO}(\omega)$ again reproduces the expected derivative-like spectral shape, however thereby overestimating the dichroism effect at the low energy resonance (see **Supplementary Fig. 3**). It can be concluded that also the magneto-optical response can be described well by using an effective medium description of the permittivity, also making these simulations significantly less computationally demanding compared to conventional layer-by-layer approaches.

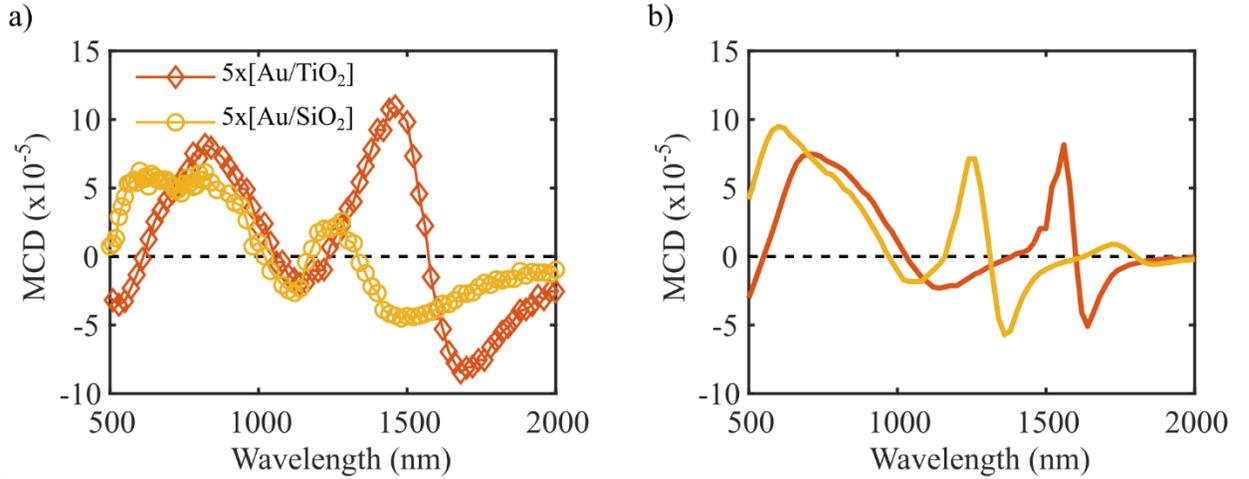

**Fig. 3. Magnetic circular dichroism in hyperbolic nanoparticles.** a) Experimental magnetic circular dichroism (MCD) spectra spectra for the hyperbolic nanoparticles made of the two Au/dielectric multilayers: for the 5x[Au/TiO$_2$] (red diamonds) and the 5x[Au/SiO$_2$] (yellow circles) samples acquired at 1.4 T applied magnetic field. b) Simulated MCD spectra using the extended effective medium approach, including a magneto-optical coupling term.

The broadband magneto-optical response of this system fundamentally originates from the anisotropy of the dielectric function, displaying type-II hyperbolicity in the visible and near-infrared spectral regions. The effect of this anisotropy is clearly observable in the numerically calculated electric near field of the resonantly excited nanoparticles, as shown in **Supplementary Fig. 4**. At the high energy resonance, the major contribution originates from the in-plane electric field (**Supplementary Fig. 4a** and **4c**), while the out-of-plane contribution dominates for the low energy resonance (**Supplementary Fig. 4b** and **4d**). These results strongly suggest a dominant role of $\varepsilon_\perp$ and $\varepsilon_\parallel$ at the low and high energy resonances, respectively. To generalize this concept, we consider a spherical particle with radius $r$ displaying either isotropic permittivity $\varepsilon_\perp$ or $\varepsilon_\parallel$ and calculate the scattering ($\sigma_{sca}$) and absorption ($\sigma_{abs}$) cross sections based on Mie theory [42,43]. The cross sections are calculated as a weighted sum of the so-called Mie coefficients $a_n$ and $b_n$, describing dipolar ($n=1$), quadrupolar ($n=2$), etc., vector spherical harmonics of electric ($a_n$) and magnetic ($b_n$) nature (details can be found in the **Methods** section) [44]. The calculated Mie efficiencies $Q_{abs/sca} = \sigma_{abs/sca}/\pi r^2$ up to $n=1$ for $r=150$ nm are presented in **Fig. 4a** and **4b** for the in-plane and out-of-plane components of $\varepsilon$, respectively. To account for the hyperbolicity of our system, we use **Eq. (1)** and **Eq. (2)** for $\varepsilon_\parallel$ and $\varepsilon_\perp$, respectively and consider a refractive index of the dielectric of 3.5 in **Eq. (2)**. In an earlier work, García-Etxarri et al. [45] showed that dielectric particles with sub-wavelength dimensions display strong and spectrally separated electric and magnetic dipolar resonances. While the in-plane-component (**Fig. 4a**) of our model system shows a metal-like response, where only an electric dipole contributes to the optical extinction, the out-of-plane component (**Fig. 4b**) exhibits a spectrally localized magnetic dipole contribution at lower energy, matching the results of García-Etxarri et al. for high-index dielectric nanoparticles. To validate our approach, we compare the isotropic Mie theory with FEM results for a sphere with anisotropic permittivity $\hat{\varepsilon} = (\varepsilon_\parallel, 0,0; 0, \varepsilon_\parallel, 0; 0,0, \varepsilon_\perp)$ (**Supplementary Fig. 6**). Indeed, as for the disks, the extinction shows high and low energy resonances which we can allocate to an in-plane electric and an out-of-plane magnetic dipole, respectively. We note that the electric near fields are spatially confined at the resonances and display a dipolar character, as can be seen in the insets of **Fig. 4c**. Consequently, even when we neglect losses in the dielectric, the system will still show disproportionately higher absorption at the low energy resonance due to the physical presence of the metal in combination with a reduced radiative coupling of the magnetic dipole. In our simple analytical model, this can be accounted for by adding a small fraction of the imaginary part of $\varepsilon_m$ to **Eq. (2)**. From a parametric study we can infer that 10% is sufficient to describe the effect, as shown in **Supplementary Fig. 5** and **Supplementary Fig. 6**. The out-of-plane electric dipole is omitted here, because the in-plane permittivity contributes dominantly at that wavelength. The sum of in-plane electric dipole and out-of-plane magnetic dipole is shown in **Fig. 4c** and agrees well with full-wave numerical results, where the complete anisotropy of the system is considered (**Supplementary Fig. 6**). For noble metals like gold, the magneto-optical coupling is predominantly determined by free charge carriers in a broad spectral range [39]. We consider the coupling of these charges to an external magnetic field in a perturbative approach [46], i.e., we introduce a magnetic-field-dependent dielectric function of the metal,

$$\varepsilon_m = \varepsilon_m^0 \pm \varepsilon_{MCD}(H). \quad (9)$$

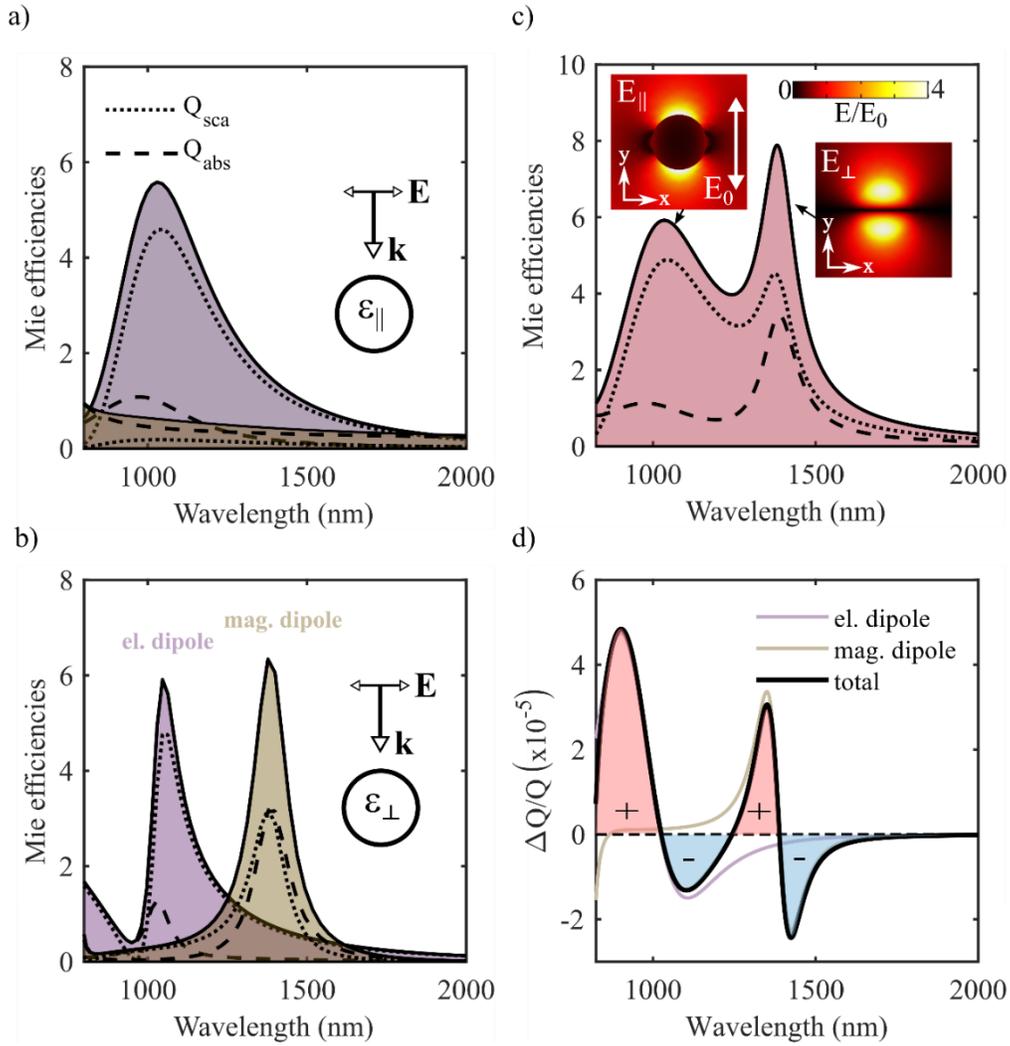

**Fig. 4. Analytical model of the magnetic circular dichroism.** Scattering (dotted lines) and absorption (dashed lines) efficiencies $Q_{sca}$ and $Q_{abs}$ for spherical nanoparticles with isotropic permittivity $\varepsilon_\parallel$ (a) and $\varepsilon_\perp$ (b). Interaction efficiencies are analytically calculated with the Mie expansion up to the first order, where magnetic (green) and electric (purple) contributions are separately evaluated. c) Sum of the in-plane electric dipole shown in (a) and the out-of-plane magnetic dipole shown in (b). The out-of-plane electric dipole is omitted, because the in-plane permittivity contributes dominantly at that wavelength. Inset are near-field plots of the resonances normalized to the incoming field $E_0$ indicated by the white arrow. d) Differential extinction spectrum $\Delta Q/Q$ calculated from the magnetically induced dichroism effect. Contributions from the electric (purple) and magnetic (green) dipole are marked by the fine solid lines. The overall spectrum (black solid line) exhibits a derivative spectral shape at the resonances, as indicated by the light blue and light red areas under the curve.

Since a change in the helicity of the incoming light is topologically equivalent to an inversion of the magnetic field direction (cf. **Fig. 2e** and **Fig. 2f**), we can also understand the MCD in terms of a magnetically induced change of the plasmonic resonance condition for a fixed polarization of the incoming light. The optical properties of the metal conduction electrons are well described by the Drude model and, in our perturbative approach, we can therefore express their magneto-optical coupling as **[18]**

$$\varepsilon_{MCD}(H) = \frac{-2\omega_c(H)\omega_p^2 \tau^3}{(\omega_c(H)^2\tau^2 - \tau^2\omega^2 + 1)^2 + 4\tau^2\omega^2}, \quad (10)$$

where $\tau$ is the electron relaxation time, $\omega_p^2 = Ne^2/\varepsilon_0 m^*$ the plasma frequency and $\omega_c(H) = eH/m^*$ the magnetic-field-dependent cyclotron frequency. In these expressions, $e$ is the electron charge, $m^*$ the effective mass and $N$ the electrons density. Inverting the applied magnetic field will thus change the cyclotron frequency of the free charge carriers, leading to a H-dependent modification of the plasmonic resonance condition, finally resulting in a spectral shift of the mode for an applied magnetic field.

Applying this approach to our system, we substitute **Eq. (9)** into **Eqs. (1)** and **(2)** and calculate the extinction efficiencies for +H and -H. Finally, due to the topological correspondence, we can express the induced magnetic circular dichroism via the change in extinction caused by inversion of the magnetic field direction as

$$\frac{\Delta Q}{Q} = \frac{Q^{(+H)} - Q^{(-H)}}{Q^{(H=0)}}, \quad (11)$$

where $Q^{(H=0)}$ is the extinction efficiency at zero magnetic field. The corresponding spectrum is shown in **Fig. 4d** (black solid line). From the Mie expansion of the extinction, we can distinguish the contributions to the dichroism coming from either the electric (fine purple line) or the magnetic (fine green line) dipole resonance, thus obtaining a derivative-like spectral shape of the differential spectrum at both resonances, which describes well the induced dichroism observed in these hyperbolic nanoparticles.

In conclusion, we demonstrated a comprehensive overview of the magneto-optics and its control in type-II hyperbolic nanoparticles by reasonably low (~1 T) external magnetic fields. Numerical simulations support well the experimental MCD in a broad spectral range and prove the viability of the effective medium theory for the description of the anisotropic permittivity of these nanoparticles. As well, numerical modelling allows addressing the magneto-optical coupling in terms of induced changes in the current density of the system, yielding an intuitive picture of the observed physical phenomenon. We established a Mie-theory-based analytical model, calculating the magneto-optical response by considering the interaction of a static magnetic field with hyperbolic dispersion-induced magnetic and electric dipolar modes of nanoparticles. Beyond fundamental insight in the magneto-photonic physics of this system, this analytical approach is general and can be expanded to other geometries and materials without time constraints related to computational approaches. Our results open an exciting path towards the development of novel nanophotonic active materials combining the benefits of sub-wavelength light confinement with the ability to tailor the magneto-optical activity at the nanoscale by exploiting hyperbolic optical dispersion.

**Methods**

**Numerical Simulations.** Simulations have been carried out with the finite elements method (FEM) implemented in the commercial COMSOL Multiphysics software. The domain was setup in 3D and truncated by perfectly matched layers to avoid back reflection from the domain walls. Left- and righthand (+ and -, respectively) circularly polarized light at different wavelengths was generated at the top of the simulation domain. The effect of a glass substrate was considered by setting the refractive index to n=1.5 in the bottom half of the simulation domain. For the layer-by-layer simulations, alternating layers of gold and the dielectric were considered. Interpolated refractive index data from Rakic et al. **[47]** for the gold and a constant refractive index for the dielectric (n(TiO$_2$)=2.0 and n(SiO$_2$)=1.75) were used to describe the linear optical properties of the system. For the homogenized structures, an effective permittivity was used, calculated according to **Eqs. (1)** and **(2)**.

**Analytical modelling.** Scattering, absorption and extinction cross sections of isotropic spheres are calculated using a customized Matlab code **[48]** following the formulation of Bohren and Huffman **[44]** for the Mie coefficients $a_n$ and $b_n$. With the size parameter $x = kr$, where $k$ is the vacuum wave number and $r$ the sphere radius, we can express the scattering and extinction efficiency as

$$Q_{sca} = \frac{2}{x^2}\sum_{n=1}^{\infty}(2n+1)\left(|a_n|^2 + |b_n|^2\right),$$

$$Q_{ext} = \frac{2}{x^2}\sum_{n=1}^{\infty}(2n+1)\text{Re}(a_n + b_n).$$

Energy conservation requires that $Q_{ext} = Q_{sca} + Q_{abs}$, thus allowing to calculate the absorption efficiency as well. Interpolated refractive index data from Rakic et al. **[47]** for the gold and a constant refractive index for the dielectric was used and the permittivity was calculated according to **Eqs. (1)** and **(2).**

**Sample fabrication.** Hyperbolic metamaterial nanoparticles and control samples (gold nano-disks) were prepared by hole mask colloidal lithography **[38]**. Microscope glass slides were cleaned with acetone and 2-propanol with 2 min sonication respectively. After deionized water (DI) washing and blow drying under $N_2$ flow, the glass wafers were ready for the multilayer deposition. For the $Au/SiO_2$ stacking layer deposition, the glass wafers were loaded into an electron beam deposition (E-beam, PVD75 Kurt J. Lesker company) chamber. One unit of the metal-dielectric bi-layer consisted of 0.5 nm Ti +10 nm Au + 0.5 nm Ti + 20 nm $SiO_2$, in which Ti served as the adhesion layer. The deposition of the bi-layer unit was repeated five times. For the $Au/TiO_2$ stacking layers, the glass wafers were loaded into an electron beam deposition chamber (Kenosistec KE 500 ET), and 0.5 nm Ti + 10 nm Au + 0.5 nm Ti layers were deposited at a rate of 0.3 Å/s. The wafer was then transferred to an atomic layer deposition chamber (ALD, FlexAL, Oxford Instruments) and $TiO_2$ was deposited using a process with titanium isopropoxide as the titanium precursor and oxygen plasma as the oxidizer. The process was repeated at 80 °C temperature for 383 cycles to produce a film with a thickness of 20nm, which was verified with ellipsometry. One unit of the $Au/TiO_2$ metal dielectric bi-layer consisted of 0.5 nm Ti + 10 nm Au + 0.5 nm Ti + 20 nm $TiO_2$. The deposition of the bi-layer unit was repeated five times. On the top of stacking bilayers, photoresist (950 PMMA A8, Micro Chem) was spin coated at 6000 rpm and soft baked at 100A °C for 1min. After $O_2$ plasma treatment (2 min, 100 W, Plasma cleaner, Gambetti), Poly(diallyldimethylammonium chloride) solution (PDDA, Mw 200,000-350,000, 20 wt. % in H2O, Sigma, three times diluted) was drop coated on the top of the PR surface and incubated for 5min to create a positively charged surface. The extra PDDA solution was washed away under flowing DI water after 5min incubation. Then negatively charged polystyrene (PS) beads (diameter 552 nm, 5 wt% water suspension, Micro Particle GmbH) were diluted 1: 3 in ethanol and drop casted on the top of the stacking bi-layers; after 30 s the excess was removed under flowing DI water and the sample was dried with $N_2$ flow. Thereby, random distributed PS beads were attached on top of the photoresist. The samples were treated with $O_2$ plasma etching in the inductively coupled plasma-reactive ion etching system (ICP-RIE, SENTECH SI500) to reduce the size of PS beads. A Gold film (40nm) was deposited with a sputter system (Sputter coater, Quorum, Q150T ES) on top of the sample to serve as an etching mask to protect the PR underneath. After removal of the PS beads with a Polydimethylsiloxane (PDMS) film, the samples were treated again by $O_2$ plasma in the ICP-RIE system to etch away the exposed PR which was no longer protected by the PS spheres and create randomly distributed holes as mask on top of the stacking bi-layers. The diameter of the holes was controlled by varying the etching time during te first $O_2$ plasma treatment performed to reduce the size of the beads. Then, 100 nm of Cr were deposited with the e-beam evaporator at a vertical cathode/target angle of incidence. Followed by liftoff of the PR in acetone, randomly distributed Cr disks on the stacking multilayer were fabricated. With the Cr disk mask, ICP-RIE etching was carried out with $CF_4$ gas flow 15 sccm, radio frequency (RF) power 200 W, ICP power 400 W, temperature 5 °C, pressure 1Pa. The etching time was adjusted according to the stacking film thickness to ensure all the extra stacking bi-layer material, except for the area under Cr mask, was removed. Then the sample was soaked in Cr etchant (Etch 18, OrganoSpezialChemie GmbH) for 2min to remove the Cr mask. Followed by DI water cleaning and drying under $N_2$ flow, the sample morphology was characterized with a scanning electron microscope (SEM, FEI Helios NanoLab 650).

**Optical and magneto-optical characterization.** Extinction spectra have been recorded on a JASCO V-670 commercial spectrophotometer in the 300 - 2700 nm spectral range. MCD spectra were collected using a home-built setup, equipped with a 300 W Xe arc lamp. The output light from a monochromator (Oriel Cornerstone 260) is linearly polarized by a Rochon prism. Circular polarization was then obtained using a photo-elastic modulator (Hinds Instruments PEM-90) able to switch at a rate of 47 kHz between right- and left-circular light polarization. Circularly polarized light is then focused on the sample, which is placed in a water-cooled electromagnet, able to reach 1.4 Tesla at room temperature. A Photomultiplier tube is used as detector for the visible range (300-850 nm), while a InGaAs diode is used for the near infrared spectral range (750-2200 nm). Light propagation was set parallel to the applied magnetic field. The dichroism signal was recorded by lock-in detection at the modulation frequency. The static signal is further modulated at 440 Hz using a mechanical chopper to filter out the residual environmental light, and subsequently analyzed by a second demodulation. The MCD signal is obtained as the ratio between the signal modulated by the PEM and the one modulated by the chopper. To avoid offset issues and spurious natural dichroism coming from the setup, the final MCD spectrum is obtained by subtraction of two spectra recorded at the same magnetic field but with opposite sign. The magnitude of the dichroism signal (ΔA) was calibrated through a standard technique using a $Fe(CN)_6^{3+}$ solution as a reference **[49]**.

**Author Contribution.** NM and FP conceived the project. JK performed numerical simulations and developed the analytical theory. AG and GP performed the magneto-optical measurements. YZ, MI and EP-V fabricated the samples. JK, AG, GP, FP and NM analyzed the data. AP, GS, AD, FDA and DB contributed to the general discussion. FP and NM supervised the work. All the authors contributed to the manuscript writing.

**Acknowledgements.** NM acknowledges support from the Luxembourg National Research Fund (Grant No. C19/MS/13624497 'ULTRON'). DB acknowledge support from the European Research Council (Grant No. 819871 'UpTEMPO'). DB and NM acknowledge support from the FEDER Program (Grant No. 2017-03-022-19 'Lux-Ultra-Fast'). AG, GP, EPV, AD and FP acknowledge the financial support of H2020

FETOPEN-2016-2017 Grant No. 737709 FEMTOTERABYTE (EC). AG, GP and FP acknowledge the financial support of Italian MIUR through PRIN 2017 (project Q-ChiSS).

## References

[1] Cortes, C. L., et al. Quantum nanophotonics using hyperbolic metamaterials. *J. Opt.* 2012, *14*(6), 063001.
[2] Lezec, H. J.; Dionne, J. A.; Atwater H. A. Negative refraction at visible frequencies. *Science* 2007, *316*, 430-432.
[3] Sreekanth, K. V.; Alapan, Y; ElKabbash, M.; Ilker, E.; Hinczewski, M.; Gurkan; U. A.; De Luca, A.; Strangi, G. Extreme sensitivity biosensing platform based on hyperbolic metamaterials. *Nat. Mater.* 2016, *15*, 621-627.
[4] Carrara, A.; Maccaferri, N.; Cerea, A.; Bozzola, A.; De Angelis, F.; Proietti Zaccaria, R.; Toma, A. Plasmon hybridization in compressible metal–insulator–metal nanocavities: an optical approach for sensing deep sub-wavelength deformation. *Adv. Opt. Mater.* 2020, *8*, 2000609.
[5] Dionne, J. A.; Sweatlock, L. A.; Atwater, H. A.; Polman, A. Planar metal plasmon waveguides: frequency-dependent dispersion, propagation, localization, and loss beyond the free electron model. *Phys. Rev. B* 2005, *72*, 075405.
[6] Caligiuri, V.; Pianelli, A.; Miscuglio, M.; Patra, A.; Maccaferri, N.; Caputo, R.; De Luca, A. Near- and mid-infrared graphene-based photonic architectures for ultrafast and low-power electro-optical switching and ultra-high resolution imaging *ACS Appl. Nano Mater.* 2020, *3*, 12218-12230.
[7] Wenyang, W.; Fan, L.; Zang, W.; Yang, X.; Zhan, P.; Chen, Z.; Wang, Z. Second harmonic generation enhancement from a nonlinear nanocrystal integrated hyperbolic metamaterial cavity. *Opt. Express* 2017, *25*, 21342-21348.
[8] Suresh, S.; Reshef, O.; Alam, M. Z.; Upham, J.; Karimi, M.; Boyd, R. W. Enhanced nonlinear optical responses of layered epsilon-near-zero metamaterials at visible frequencies. *ACS Photonics* 2021, *8*, 125-129.
[9] Kuttruff, J.; Garoli, D. ; Allerbeck, J.; Krahne, R.;  De Luca, A.; Brida, D.; Caligiuri, V.; Maccaferri, N. Ultrafast all-optical switching enabled by epsilon-near-zero-tailored absorption in metal-insulator nanocavities. *Commun. Phys.* 2020, *3*, 114.
[10] Maccaferri, N.; Zilli, A.; Isoniemi, T.; Ghirardini, L.; Iarossi, M.; Finazzi, M.; Celebrano, M.; De Angelis, F. Enhanced nonlinear emission from single multilayered metal–dielectric nanocavities resonating in the near-infrared *ACS Photonics* 2021.
[11] Indukuri, S. R. K. C.; Bar-David, J.; Mazurski, N.; Levy, U. Ultrasmall mode volume hyperbolic nanocavities for enhanced light–matter interaction at the nanoscale. *ACS Nano* 2019, *13*, 11770-11780.
[12] Lu, D.; Qian, H.; Wang, K.; Shen, H.; Wei, F.; Jiang, Y.; Fullerton, E. E.; Yu, P. K. L.; Liu, Z. Nanostructuring multilayer hyperbolic metamaterials for ultrafast and bright green InGaN quantum wells. *Adv. Mater.* 2018, *30*, 1706411.
[13] Indukuri, S. R. K. C.; Frydendahl, C.; Bar-David, J.; Mazurski, N.; Levy, U. WS2 monolayers coupled to hyperbolic metamaterial nanoantennas: broad implications for light–matter-interaction applications. *ACS Appl. Nano Mater.* 2020, *3*, 10226-10233.
[14] Lu, D.; Kan, J. J.; Fullerton, E. E.; Liu, Z. Enhancing spontaneous emission rates of molecules using nanopatterned multilayer hyperbolic metamaterials. *Nat. Nanotechnol.* 2014, *9*, 48-53.
[15] Caligiuri, V.; Dhama, R.; Sreekanth, K. V.; Strangi, G.; De Luca, A. Dielectric singularity in hyperbolic metamaterials: the inversion point of coexisting anisotropies. *Sci. Rep.* 2016, *6*, 20002.
[16] Wang, X.; Wang, H.; Jian, J.; Rutherford, B. X.; Gao, X.; Xu, X.; Wang, H. Metal-free oxide-nitride heterostructure as a tunable hyperbolic metamaterial platform. *Nano Lett.* 2020, *20*(9), 6614-6622.
[17] López-Ortega, A.; Zapata-Herrera, M.; Maccaferri, N.; Pancaldi, M.; Garcia, M.; Chuvilin, A.; Vavassori, P. Enhanced magnetic modulation of light polarization exploiting hybridization with multipolar dark plasmons in magnetoplasmonic nanocavities. *Light Sci. Appl.* 2020, *9(1)*, 1-14.

[18] Armelles, G.; Cebollada, A.; García-Martín, A.; González, M. U. Magnetoplasmonics: combining magnetic and plasmonic functionalities. *Adv. Opt. Mater.* **2013**, *1*(1), 10-35.
[19] Maccaferri, N. Coupling phenomena and collective effects in resonant meta-atoms supporting both plasmonic and (opto-)magnetic functionalities: an overview on properties and applications. *J. Opt. Soc. Am. B* **2019**, *36*, E112-E131.
[20] Maccaferri, N.; Zubritskaya, I.; Razdolski, I.; Chioar, I.-A.; Belotelov, V.; Kapaklis, V.; Oppeneer, P. M.; Dmitriev, A. Nanoscale magnetophotonics. *J. Appl. Phys.* **2020**, *127*, 080903.
[21] Pineider, F.; Campo, G.; Bonanni, V.; de Julián Fernández, C.; Mattei, G.; Caneschi, A.; Sangregorio, C. Circular magnetoplasmonic modes in gold nanoparticles. *Nano Lett.* **2013**, *13*(10), 4785-4789.
[22] Gabbani, A.; Fantechi, E.; Petrucci, G.; Campo, G.; de Julián Fernández, C.; Ghigna, P.; Pineider, F. Dielectric Effects in FeO x-Coated Au Nanoparticles Boost the Magnetoplasmonic Response: Implications for Active Plasmonic Devices. *ACS Appl. Nano Mater.* **2021**, DOI: 10.1021/acsanm.0c02588
[23] Maccaferri, N.; Gregorczyk, K. E.; de Oliveira, T. V. A. G.; Kataja, M.; van Dijken, S.; Pirzadeh, Z.; Dmitriev, A.; Åkerman, J.; Knez, M.; Vavassori, P. Ultrasensitive and label-free molecular-level detection enabled by light phase control in magnetoplasmonic nanoantennas. *Nat. Commun.* **2015**, *6*, 6150.
[24] Diaz-Valencia, B. F.; Mejía-Salazar, J. R.; Oliveira Jr., O. N.; Porras-Montenegro, N.; Albella, P. Enhanced transverse magneto-optical Kerr effect in magnetoplasmonic crystals for the design of highly sensitive plasmonic (bio)sensing platforms. *ACS Omega* **2017**, *2*, 7682-7685.
[25] Tran, V. T.; Kim, J.; Tufa, L. T.; Oh, S.; Kwon, J.; Lee, J. Magnetoplasmonic nanomaterials for biosensing/imaging and in vitro/in vivo biousability. *Anal. Chem.* **2018**, *90*, 225-239.
[26] Ignatyeva, D. O.; Kapralov, P. O.; Knyazev, G. A.; Sekatskii, S. K.; Dietler, G.; Nur-E-Alam, M.; Vasiliev, M.; Alameh, K.; Belotelov, V. I. High-Q surface modes in photonic crystal/iron garnet film heterostructures for sensor applications. *JETP Letters* **2016**, *104*, 679-684.
[27] Manera, M. G.; Colombelli, A.; Taurino, A.; García-Martín, A.; Rella, A. Magneto-Optical properties of noble-metal nanostructures: functional nanomaterials for bio sensing. *Sci. Rep.* **2018**, *8*, 12640.
[28] Ignatyeva, D. O.; Davies, C. S.; Sylgacheva, D. A.; Tsukamoto, A.; Yoshikawa, H.; Kapralov, P. O.; Kirilyuk, A.; Belotelov, V. I.; Kimel, A. V. Plasmonic layer-selective all-optical switching of magnetization with nanometer resolution. *Nat. Commun.* **2019**, *10*, 4786.
[29] Novikov, I. A.; Kiryanov, M. A.; Nurgalieva, P. K.; Frolov, A.Yu.; Popov, V. V.; Dolgova, T. V.; Fedyanin, A. A. Ultrafast magneto-optics in nickel magnetoplasmonic crystals. *Nano Lett.* **2020**, *20*, 8615-8619.
[30] Lodewijks, K.; Maccaferri, N.; Pakizeh, T.; Dumas, R. K.; Zubritskaya, I.; Åkerman, J.; Vavassori, P.; Dmitriev, A. Magnetoplasmonic design rules for active magneto-optics. *Nano Lett.* **2014**, *14*, 7207–7214.
[31] Chin, J. Y.; Steinle, T.; Wehlus, T.; Dregely, D.; Weiss, T.; Belotelov, V. I.; Stritzker, B.; Giessen, H. Nonreciprocal plasmonics enables giant enhancement of thin-film Faraday rotation. *Nat. Commun.* **2013**, *4*, 1599.
[32] Floess, D.; Hentschel, M.; Weiss, T.; Habermeier, H.-U.; Jiao, J.; Tikhodeev, S. G.; Giessen, H. Plasmonic analog of electromagnetically induced absorption leads to giant thin film faraday rotation of 14°. *Phys. Rev. X* **2017**, *7*, 021048.
[33] Floess, D.; Giessen H. Nonreciprocal hybrid magnetoplasmonics. *Rep. Prog. Phys.* **2018**, *81*,116401.
[34] Fan, B.; Nasir, M. E.; Nicholls, L. H.; Zayats, A. V.; Podolskiy, V. A. Magneto-Optical Metamaterials: Nonreciprocal Transmission and Faraday Effect Enhancement. *Adv. Opt. Mater.* **2019**, *7*(14), 1801420.
[35] Song, J.; Cheng, Q.; Lu, L.; Li, B.; Zhou, K.; Zhang, B.; Zhou, X. Magnetically tunable near-field radiative heat transfer in hyperbolic metamaterials. *Phys. Rev. Applied* **2020**, *13*(2), 024054.
[36] Maccaferri, N.; Zhao, Y.; Isoniemi, T.; Iarossi, M.; Parracino, A.; Strangi, G., De Angelis, F. Hyperbolic meta-antennas enable full control of scattering and absorption of light. *Nano Lett.* **2019**, *19*, 1851-1859.
[37] Isoniemi, T.; Maccaferri, N.; Ramasse, Q. M.; Strangi, G.; De Angelis, F. Electron energy loss spectroscopy of bright and dark modes in hyperbolic metamaterial nanostructures. *Adv. Opt. Mater.* **2020**, *8*, 2000277.
[38] Fredriksson, H.; Alaverdyan, Y.; Dmitriev, A.; Langhammer, C.; Sutherland, D. S.; Zäch, M.; Kasemo, B. Hole–mask colloidal lithography. *Adv. Mater.* **2007**, *19*(23), 4297-4302.
[39] Sepúlveda, B.; González-Díaz, J. B.; García-Martín, A.; Lechuga, L. M.; Armelles, G. Plasmon-induced magneto-optical activity in nanosized gold disks. *Phys. Rev. Lett.* **2010**, *104*(14), 147401.

[40] Maccaferri, N.; Berger, A.; Bonetti, S.; Bonanni, V.; Kataja, M.; Qin, Q. H.; Vavassori, P. Tuning the magneto-optical response of nanosize ferromagnetic Ni disks using the phase of localized plasmons. *Phys. Rev. Lett.* **2013**, *111*(16), 167401.

[41] Han, B.; Gao, X.; Lv, J.; Tang, Z. Magnetic circular dichroism in nanomaterials: New opportunity in understanding and modulation of excitonic and plasmonic resonances. *Advanced Materials* **2020**, *32*(41), 1801491.

[42] Mie, G. "Beiträge zur optik trüber medien, speziell kolloidaler metallösungen." *Ann. Phys.* **1908**, *330*, 377-445.

[43] Schmidt, M. K.; Esteban, R.; Sáenz, J.; Suárez-Lacalle, I.; Mackowski, S.; Aizpurua, J. Dielectric antennas-a suitable platform for controlling magnetic dipolar emission. *Opt. Express* **2012**, *20*(13), 13636-13650.

[44] Bohren, Craig F., and Donald R. Huffman. Absorption and scattering of light by small particles. John Wiley & Sons **2008**.

[45] García-Etxarri, A.; Gómez-Medina, R.; Froufe-Pérez, L. S.; López, C.; Chantada, L.; Scheffold, F.; Sáenz, J. J. Strong magnetic response of submicron silicon particles in the infrared. *Opt. Express* **2011**, *19*(6), 4815-4826.

[46] Yu, G.; Kornev, K. G. Plasmon enhanced direct and inverse Faraday effects in non-magnetic nanocomposites. *J. Opt. Soc. Am. B* **2010**, *27*, 2165-2173.

[47] Rakić, A.; Djurišić, A.; Elazar, J.; Majewski, M. Optical properties of metallic films for vertical-cavity optoelectronic devices. *Appl. Opt.* **1998**, *37*, 5271.

[48] Matzler, C. MATLAB functions for Mie scattering and absorption. IAP Res. Rep. No. 2002-11, **2002**.

[49] Mason, W. R. A practical guide to magnetic circular dichroism spectroscopy. Wiley-Interscience **2007**.

## Supplementary Information

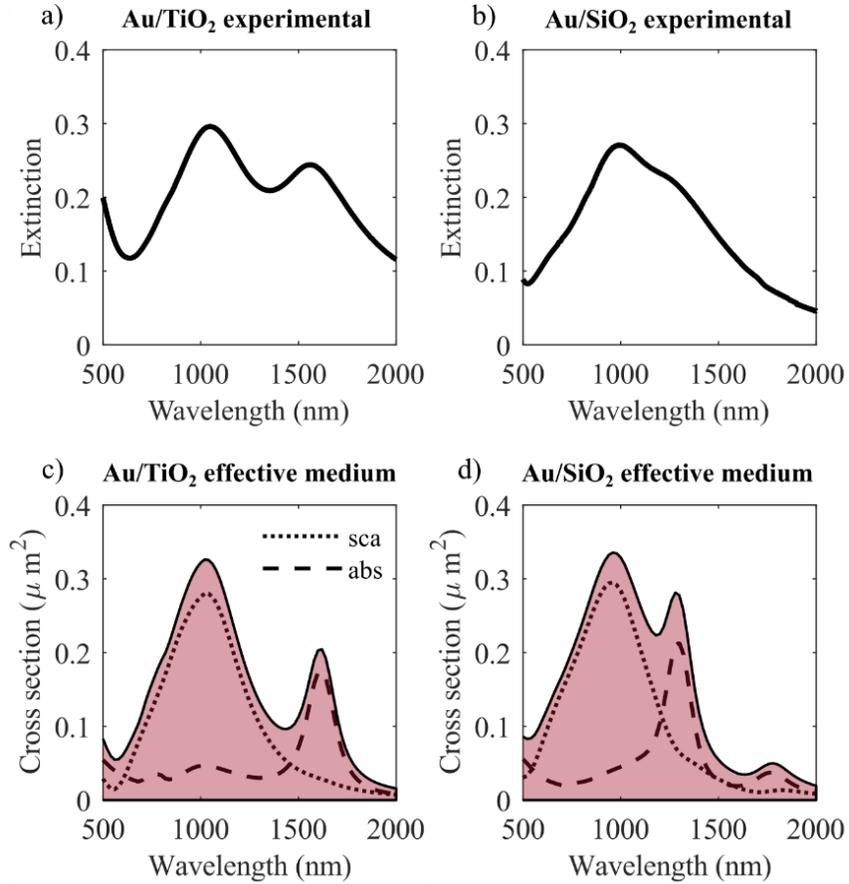

SUPPLEMENTARY FIG. 1. Experimental extinction (a and b) and simulated scattering (dotted lines) and absorption (dashed lines) cross sections (c and d) for the Au/TiO$_2$ and the Au/SiO$_2$ samples as a function of the wavelength of the incoming light.

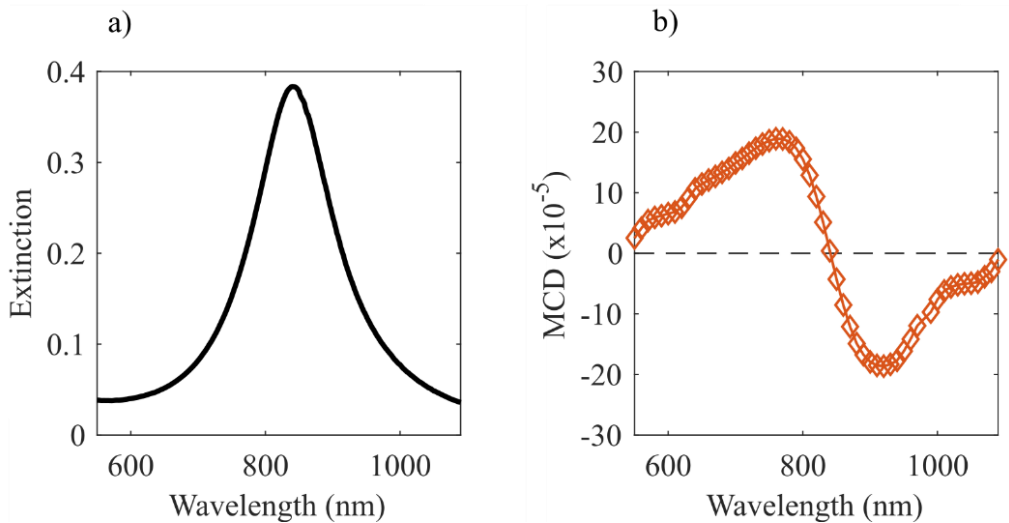

SUPPLEMENTARY FIG. 2. Experimental extinction (a) and MCD (b) spectra of plain Au nanodisks.

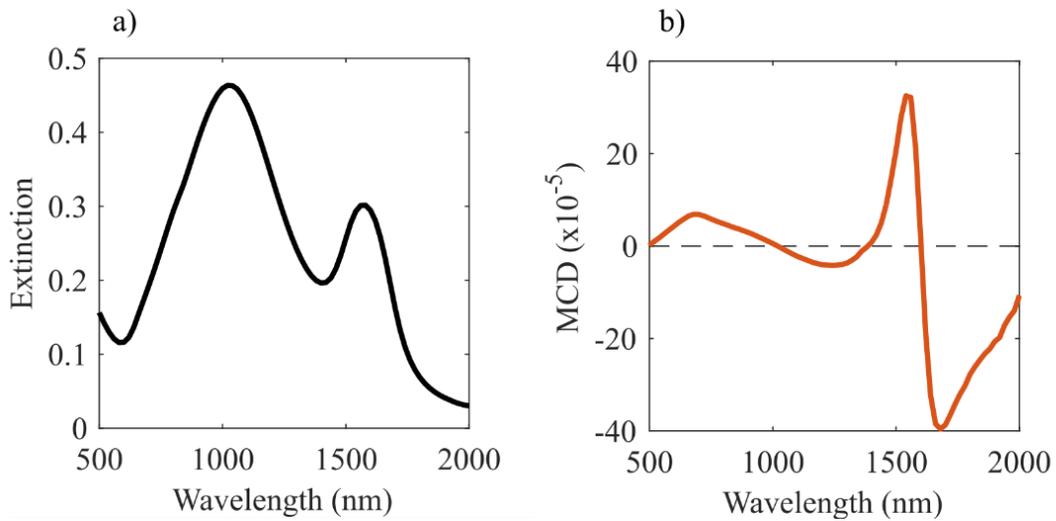

SUPPLEMENTARY FIG. 3. Numerical extinction (a) and MCD (b) spectra of the hybrid Au/TiO$_2$ structure using a frequency dependent off-diagonal permittivity $\varepsilon_{MO}(\omega) = i\varepsilon(\omega)\frac{\omega_c}{\omega+i\gamma}$, where $\varepsilon(\omega)$ is the diagonal permittivity, $\omega_c$ is the cyclotron frequency and $\gamma$ is the electronic relaxation constant.

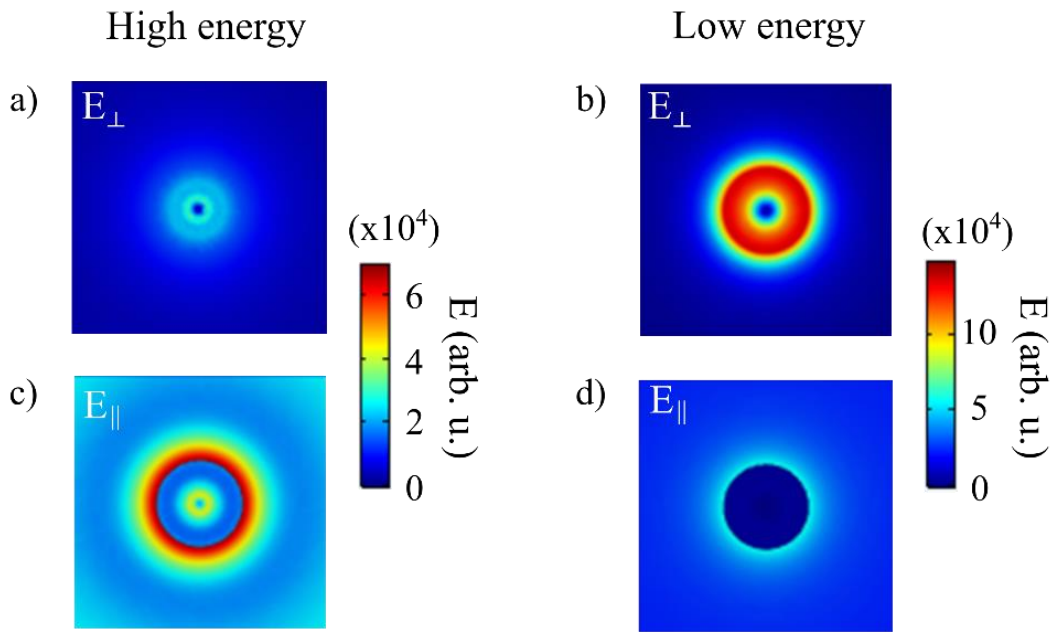

SUPPLEMENTARY FIG. 4. Electric near-field plots of the hyperbolic Au/TiO2 nanodisk at the high (a) and c)) and low (b) and d)) energy resonance for the in- and out-of-plane component of the field, respectively. X-Y cuts are taken in the middle of the structure. While the in-plane component dominates at the high energy resonance, the out-of-plane component majorly contributes at the low energy resonance.

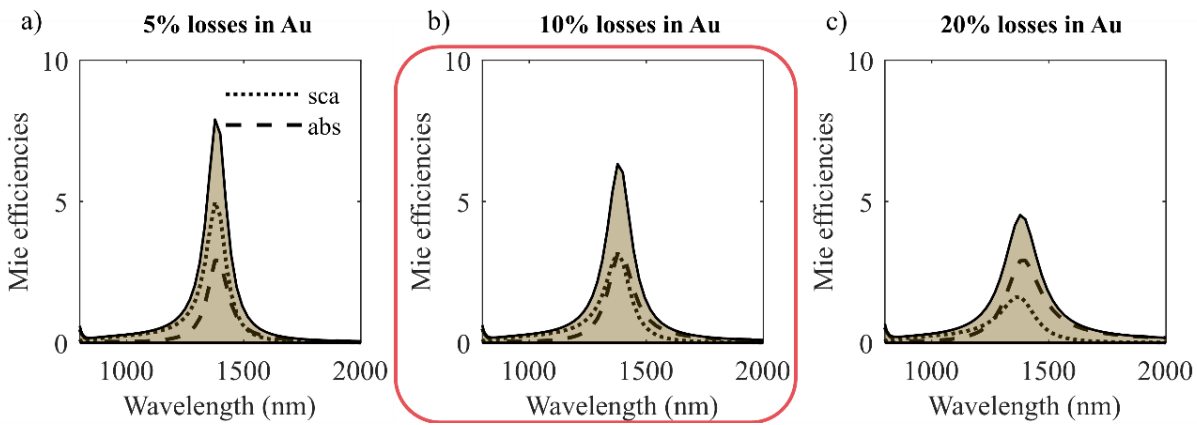

SUPPLEMENTARY FIG. 5. Calculated scattering (dotted lines) and absorption (dashed lines) efficiencies of the out-of-plane magnetic dipole term in the Mie expansion of the optical extinction. A small fraction (5% (a), 10% (b) and 20% (c)) of the imaginary part of the metal permittivity is added to the out-of-plane dielectric function to account for non-radiative losses in gold. A fraction of 10 % (marked by the red box) yields equal contribution of scattering and absorption, representing well the numerical results for real anisotropic particles at the low energy resonance shown in Supplementary Fig. 5.

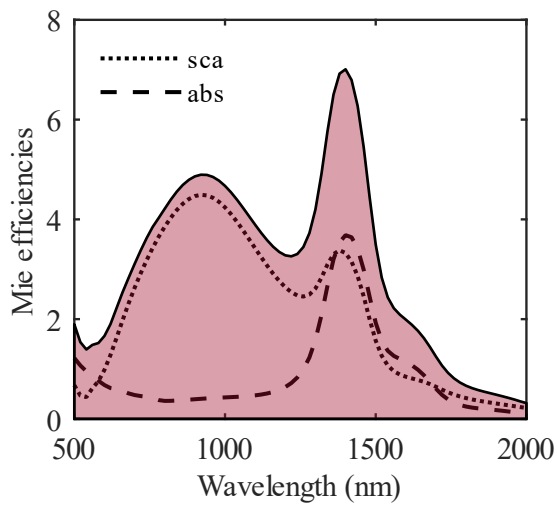

SUPPLEMENTARY FIG. 6. Numerically calculated scattering (dotted line) and absorption (dashed line) efficiencies of a sphere with $r = 150$ nm considering anisotropic permittivity $\hat{\varepsilon} = (\varepsilon_\parallel, 0, 0; 0, \varepsilon_\parallel, 0; 0, 0, \varepsilon_\perp)$. While scattering dominates at the high energy resonance, we find an equal contribution of scattering and absorption at the low energy resonance.